\documentclass[reprint,amsmath,amssymb,aps,nofootinbib,longbibliography,superscriptaddress]{revtex4-1}
\usepackage{graphicx}
\usepackage{dcolumn}
\usepackage{bm}
\usepackage{bbold}
\usepackage{natbib}
\usepackage{hyperref} %needs to go before hypersetup
\hypersetup{colorlinks=true,linkcolor=blue,filecolor=magenta,urlcolor=cyan}

\usepackage{siunitx}
\usepackage{commath}
\usepackage{multirow}
\usepackage{hhline}
\DeclareMathAlphabet\mathbfcal{OMS}{cmsy}{b}{n}
\usepackage{color}

\begin{document}

\title{Exploiting Faraday Rotation to Jam Quantum Key Distribution via Polarized Photons}
\author{Maximilian Daschner}\email{daschner@mit.edu}
\affiliation{ETH Zurich, 8093 Switzerland}
\affiliation{Massachusetts Institute of Technology, Cambridge, Massachusetts 02139, USA}
\author{David I. Kaiser}\email{dikaiser@mit.edu}
\author{Joseph A. Formaggio}\email{josephf@mit.edu}
\affiliation{Massachusetts Institute of Technology, Cambridge, Massachusetts 02139, USA}
\date{\today}

\begin{abstract}
Quantum key distribution (QKD) involving polarized photons could be vulnerable to a jamming (or denial-of-service) attack, in which a third party applies an external magnetic field to rotate the plane of polarization of photons headed toward one of the two intended recipients. Sufficiently large Faraday rotation of one of the polarized beams would prevent Alice and Bob from establishing a secure quantum channel. We investigate requirements to induce such rotation both for free-space transmission and for transmission via optical fiber, and find reasonable ranges of parameters in which a jamming attack could be successful against fiber-based QKD, even for systems that implement automated recalibration for polarization-frame alignment. The jamming attack could be applied selectively and indefinitely by an adversary without revealing her presence, and could be further combined with various eavesdropping attacks to yield unauthorized information.
\end{abstract}
\maketitle
\section{Introduction}

One of the most remarkable technologies associated with quantum entanglement is quantum encryption \cite{Nielsen,Gisin,AlleaumeSurvey,Shenoy,PirandolaReview}. In the popular Ekert91 protocol \cite{ekert}, a source creates pairs of entangled particles; one member of each pair goes to Alice and Bob, respectively. By performing measurements at their respective stations on a series of entangled particles and sharing some information on a classical channel about their choices of measurement bases (but {\it not} sharing their measurement outcomes), Alice and Bob can establish a %quantum 
secret key with which to encrypt and share messages. They can further establish the security of their quantum channel by subjecting a subset of their entangled particles to a Bell test, by measuring (for example) the Clauser-Horne-Shimony-Holt parameter $S$ \cite{clauser}.

In recent years, quantum key distribution (QKD) has advanced to robust, in-the-field demonstrations. Secure 
%quantum 
keys have been distributed using photons through optical fibers as well as via free-space transmission. Perhaps the most dramatic demonstrations to date have involved quantum key distribution across thousands of kilometers using the {\it Micius} satellite, launched by the Chinese Academy of Sciences in 2016 and now in low-Earth orbit at an average altitude of 500 km \cite{LiaoSatellite,YinSatelliteQKD,MiciusQKD}. 
Shorter-distance quantum key distribution is typically accomplished via networks of optical fibers, over distances ranging from a few kilometers to hundreds of kilometers \cite{PanReviewLargeScale}.

In principle, the security of QKD derives from fundamental properties of quantum mechanics, such as entanglement, the uncertainty principle, and the no-cloning theorem \cite{Nielsen,Gisin,AlleaumeSurvey,Shenoy,PirandolaReview}. In practice, physicists and engineers have identified a series of potential vulnerabilities that could enable an adversary to eavesdrop on a quantum channel and gain information not intended for them \cite{Gisin,Scarani,PirandolaReview,ScaraniBlackPaper,AlleaumeSurvey,DiamantiReview,PanReviewRealisticQKD}. 

To date the main concern in studies of QKD has been to find ways to prevent an eavesdropper from gaining unauthorized information about the secret key. Yet quantum channels are also vulnerable to jamming (or denial-of-service) attacks, in which the intended recipients would not be able to verify the security of the quantum channel and hence would refrain from using it to communicate \cite{AlleaumeSurvey,Schartner,HuguesSalas2,LiDOS,CaiPingPong}. It has long been recognized that an adversary could block a given channel by simply cutting the optical fiber connecting Alice's and Bob's stations. Nevertheless, as reliance upon extended QKD networks becomes more common, it is worthwhile considering a range of methods by which such denial-of-service attacks might be accomplished. In particular, if an adversary simply cut the optical fiber with which Alice and Bob had established a secure quantum channel, she would immediately alert Alice and Bob that their system was under attack, and would forfeit any further control of their system. 
%We argue here that an adversary could gain certain advantages by using a method to jam QKD that is more subtle than cutting the fiber.

%We argue here that an adversary could gain certain advantages by using a method more subtle than cutting the optical fiber. In particular, if an adversary simply cut the optical fiber with which Alice and Bob had established a secure quantum channel, she would immediately alert Alice and Bob that their system was under attack.

In this paper we consider a different approach, in which an adversary could {\it selectively} jam QKD indefinitely, without revealing her presence. The adversary could accomplish this jamming attack without gaining direct access to the optical fiber or to Alice's and Bob's instrumentation, and hence the attack could be perpetrated inconspicuously, at a significant distance from Alice's and Bob's stations. The adversary would maintain complete control over when and for how long a given jamming attack would occur. 

In addition, the selective jamming attack that we identify can be used in combination with other attacks that could (potentially) yield valuable information to an eavesdropper. For example, an adversary could use our jamming attack to induce Alice and/or Bob to undertake a new calibration of their devices at a time to be determined by the eavesdropper, and then exploit a ``device calibration" attack \cite{JainDeviceCalibration} to obtain unauthorized information. The jamming attack that we describe here would also likely accentuate detector-efficiency mismatches at Alice's and/or Bob's detectors, which could be exploited as part of a time-shift attack or a related eavesdropping attack that depends upon such efficiency mismatches \cite{timeshift1,timeshift2,timeshiftaltern,PirandolaReview,FungDEM,LydersenDEM,FeiDEM}. Such a two-part strategy would not be available to an attacker who simply cut the optical fiber connecting Alice's and Bob's stations.

Our approach identifies a potential vulnerability for QKD systems that use polarized photons. By applying an external magnetic field to some limited region of space through which the photons intended for Alice (or Bob) must pass, an adversary could induce a rotation of the photons' polarization via the Faraday effect. A sufficiently strong magnetic field would reduce the degree of Bell violation that Alice and Bob would measure, to the point that they would not be able to verify that they shared a trustworthy quantum channel. Hence they would not be able to establish a secure quantum key. The same type of jamming attack would be effective against other QKD protocols involving polarized photons, such as variants of the BB84 protocol \cite{bennett}, which do not involve explicit Bell-CHSH measurements, since the Faraday rotation would increase the quantum bit error rate (QBER).

In Section \ref{sec:BeamRotation}, we briefly derive the effect that a rotation of the plane of polarization of one of the polarized photon beams would have on Alice's and Bob's measurement of the CHSH parameter $S$, from which we may evaluate the magnitude of a rotation angle $\alpha$ required for a jamming attack. In Section \ref{sec:Verdet} we consider applications of such a jamming attack to free-space QKD and to distribution via optical fiber. As we will see, the scenario we describe would not likely be effective against free-space QKD involving polarized photons, but could have a significant impact on systems that depend upon optical fiber. Concluding remarks follow in Section \ref{sec:Conclusions}.

\section{Beam Rotation and Bell Tests}
\label{sec:BeamRotation}

In order to confirm the security of a quantum channel, Alice and Bob may perform a Bell test, using a portion of the entangled photons that have been distributed from the source. We assume that the source emits pairs of polarization-entangled photons in a maximally entangled state such as $\vert \Phi^\pm \rangle = \{ \vert H\rangle_A \otimes \vert H\rangle_B  \pm \vert V\rangle_A \otimes \vert V\rangle_B  \} / \sqrt{2}$, where subscripts $A, B$ indicate photons directed toward Alice and Bob, respectively. Alice and Bob can each measure polarization in one of two bases, represented by the spatial unit vectors ${\bf a}$ and ${\bf a}'$ for Alice and ${\bf b}$ and ${\bf b}'$ for Bob; we label their measurement outcomes ${\cal A, B} \in \{ +1, -1 \}$. Then $p ({\cal A = B} \vert {\bf a}, {\bf b})$ is the conditional probability that Alice and Bob will obtain the same measurement outcome given the joint settings $({\bf a}, {\bf b})$, in terms of which we may define the correlation function $E ({\bf a}, {\bf b}) = 2p ({\cal A = B} \vert {\bf a}, {\bf b}) - 1$. The CHSH parameter $S$ takes the form \cite{clauser}
%%%%%
\begin{equation}
    S \equiv \vert E ({\bf a}, {\bf b}) - E ({\bf a}, {\bf b}') + E ({\bf a}' , {\bf b}) + E ({\bf a}', {\bf b}') \vert .
    \label{Sdef}
\end{equation}
Measurements on particles that are not entangled obey the Bell-CHSH inequality, $S \leq 2$, whereas measurements on entangled particles may yield $S > 2$ \cite{clauser}. In particular, the maximum violation of the Bell-CHSH inequality consistent with quantum mechanics is the Tsirelson bound, $S_{\rm max} = 2 \sqrt{2}$ \cite{Cirelson}. For measurements of polarized photons in the state $\vert \Phi^\pm \rangle$, quantum mechanics predicts that the correlation functions should behave as $E ({\bf a}, {\bf b}) \rightarrow  \cos (2 \theta_{{\bf a} {\bf b}})$, where $\cos \theta_{\bf a b} \equiv {\bf a} \cdot {\bf b}$. If we assume that the entangled photons travel along ${\bf z}$ and Alice and Bob measure their polarizations within the ${\bf x} - {\bf y}$ plane, then the Tsirelson bound corresponds to the choice of bases $({\bf a}, {\bf a}') = (0^\circ, 45^\circ)$ and $({\bf b}, {\bf b}' ) = (22.5^\circ , 67.5^\circ)$.

If Alice or Bob measure their photons in some rotated basis, then the value of $S$ that they measure will decrease from its maximum value. Consider the case in which Alice measures her photons in a basis rotated by an angle $\alpha$ in the ${\bf x} - {\bf y}$ plane: $({\bf a}, {\bf a}') \rightarrow (\tilde{\bf a} , \tilde{\bf a}') = (0^\circ + \alpha , 45^\circ + \alpha)$.
%, with ${\bf a} \cdot \tilde{\bf a} = {\bf a}' \cdot \tilde{\bf a}' = \cos \alpha$. 
Then it is straightforward to demonstrate that the Bell-CHSH violation for the state $\vert \Phi^\pm \rangle$ measured in the bases $(\tilde{\bf a} , \tilde{\bf a}')$ and $({\bf b}, {\bf b}')$ will shift to $S \rightarrow \tilde{S}$, with
%%%%%
\begin{equation}
 \tilde{S} = S \, \vert \cos (2 \alpha ) \vert .
\label{Stilde}    
\end{equation}
The same shift $S \rightarrow \tilde{S}$ will result if we keep Alice's measurement bases $({\bf a}, {\bf a}')$ unchanged but rotate the plane of polarization of the photons directed towards her by an angle $\alpha$. In that case $\vert H\rangle_A \rightarrow \vert \tilde{H} \rangle_A =\cos \alpha \vert H \rangle_A + \sin \alpha \vert V \rangle_A$ and $\vert V\rangle_A \rightarrow \vert \tilde{V} \rangle_A = - \sin \alpha \vert H \rangle_A + \cos \alpha \vert V \rangle_A$, and the entangled state shifts to $\vert \Phi^\pm \rangle \rightarrow \vert \tilde{\Phi}^\pm \rangle = \{ \vert \tilde{H} \rangle_A \otimes \vert H \rangle_B \pm \vert \tilde{V} \rangle_A \otimes \vert V \rangle_B \} / \sqrt{2}$. Upon noting that for the states $\vert \Psi^\pm \rangle = \{ \vert H \rangle_A \otimes \vert V \rangle_B \pm \vert V \rangle_A \otimes \vert H \rangle_B \} / \sqrt{2}$ the quantum-mechanical correlation function behaves as $E ({\bf a}, {\bf b} ) = - \cos (2 \theta_{\bf ab} )$, we again find the predicted shift in the CHSH parameter as given in Eq.~(\ref{Stilde}).

In a real test, Alice and Bob will typically measure a value $2 < S_{\rm meas} < 2\sqrt{2}$. The visibility fraction $S_{\rm meas} / 2 \sqrt{2}$ will fall below one due to imperfections in the state preparation, such that the photons that Alice and Bob receive are not in a maximally entangled state; misalignments of the bases from the ideal settings; finite-aperture effects; limited detector efficiencies, and so on \cite{Brunner}. Nonetheless, Alice and Bob can have confidence in the security of their quantum channel if their measured value $S_{\rm meas} \pm \sigma$ (taking into account systematic errors, quantified by the standard deviation $\sigma$) exceeds the Bell-CHSH limit of $2$ to high statistical significance.

\begin{figure}[ht]
\centering
  \includegraphics[scale=0.24]{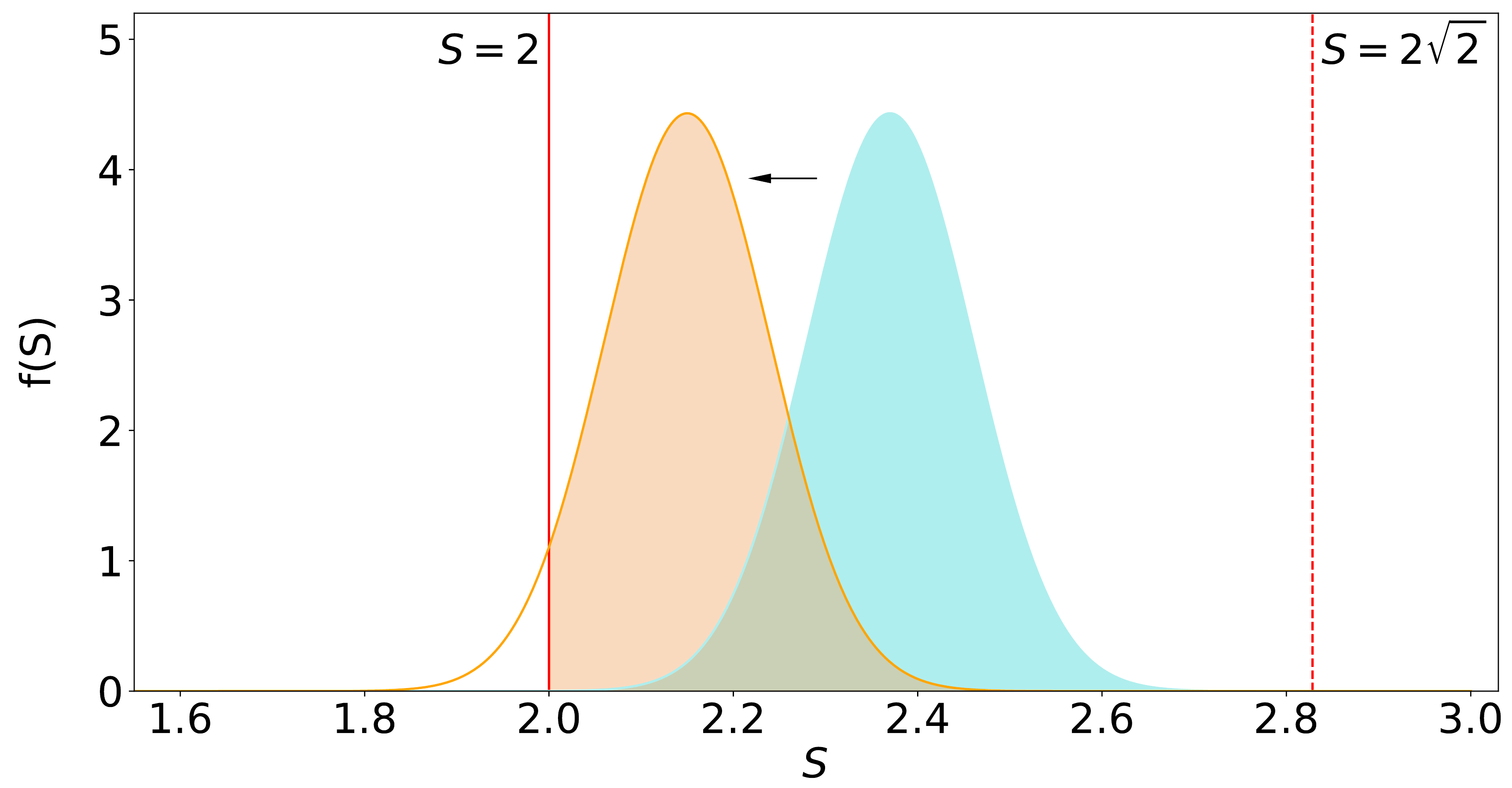}
%  \caption{Gauss Curve Shift}
\caption{Normalized probability density functions $f (S_{\rm meas} \vert \mu, \sigma^2)$ for expected values to be measured of the CHSH parameter $S$, assuming Gaussian statistics (with mean $\mu$ and standard deviation $\sigma$). A rotation of the polarization of the photons directed toward Alice will shift the mean measured value of $S$ toward lower values. In this case, a rotation of the polarization of Alice's photons by $\alpha = 12.4^\circ$ shifts the measured value $S_{\rm meas} = 2.37 \pm 0.09$ (blue curve) to $\tilde{S}_{\rm meas} = 2.15\pm 0.09$ (orange curve). Also shown are the Bell-CHSH values $S = 2$ and the Tsirelson bound $S = 2 \sqrt{2}$. Because the correlation functions $E ({\bf a}, {\bf b})$ satisfy $\vert E ({\bf a}, {\bf b} ) \vert \leq 1$, the maximum (algebraic) value for $S$ is 4.}
  \label{fig:Gauss}
\end{figure}

If --- unbeknownst to Alice --- the polarization of her photons has been rotated by a sufficiently large angle $\alpha$ en route to her, after their emission from the source, then she and Bob will {\it not} be able to verify with sufficient confidence that they share a secure quantum channel. We make the conservative assumption that the rotation of the polarization of Alice's photons does not increase the systematic errors $\sigma$ for Alice's and Bob's measurements of $S$, since nothing has been done to affect the instrumentation at either receiving station: both sides still operate with the same detector efficiencies, dark count rates, and the like. If we further assume that Alice and Bob perform measurements on a sufficiently large sample of photons that they may assume Gaussian statistics, then the effect of the rotated polarization will be to shift the distribution of values $S_{\rm meas}$ that they are likely to find, with a mean value closer to the Bell-CHSH limit of $S = 2$, as shown in Fig.~\ref{fig:Gauss}.

In Fig.~\ref{fig:Gauss} we illustrate the effects of an induced shift $S_{\rm meas} \rightarrow \tilde{S}_{\rm meas}$. In a recent experimental test of the Bell-CHSH inequality using polarization-entangled photons from the satellite {\it Micius}, receiving stations on the ground measured $S_{\rm meas} = 2.37 \pm 0.09$ \cite{yin1}. If the polarization of Alice's photons had been rotated by $\alpha = 0.216 \> {\rm rad} = 12.4^\circ$ en route, while all other details of the experimental test remained unchanged, then from Eq.~(\ref{Stilde}) we would expect a shift $S_{\rm meas} \rightarrow \tilde{S}_{\rm meas} = 2.15 \pm 0.09$. In that case, what had originally been a violation of the Bell-CHSH inequality by more than 4 standard deviations would be reduced to a violation by just $1.7$ standard deviations. Assuming Gaussian statistics, this would correspond to a shift in the probability with which Alice and Bob would conclude that their quantum channel was {\it insecure} from $p = 2.0 \times 10^{-5}$ to $p = 0.05$. Given such a relatively large $p$ value in the latter case, Alice and Bob would not be justified in trusting their quantum channel.

For other QKD protocols involving polarized photons, such as the BB84 protocol and its variants \cite{bennett}, Alice and Bob need not explicitly perform a Bell-CHSH test. Nonetheless, such protocols would also be affected if, for some reason, the plane of polarization of the photons headed toward Alice were rotated by an angle $\alpha.$ In particular, such rotation would increase the quantum bit error rate (QBER), potentially endangering the ability of Alice and Bob to establish a secure quantum channel. In this case, a photon in a particular quantum state, such as $\vert \tilde{H}\rangle_A$, would be misidentified by Alice's detector as being $\vert V \rangle_A$ a fraction $(1 - \cos^2 \alpha)$ of times (and likewise $\vert \tilde{V} \rangle_A$ for $\vert H \rangle_A$). The additional QBER would be about $7.6\%$ for $\alpha = 0.279 \> {\rm rad} = 16.0^\circ$. Given typical QBER for QKD systems under nominal operating conditions of $2 - 3\%$ \cite{ZeilingerTokyo,JenneweinAlignment,PanReviewRealisticQKD}, the additional error rate induced by the rotation angle $\alpha$ would push the total QBER too high for reliable use.

\section{Jamming QKD Systems}
\label{sec:Verdet}

How might an adversary (whom we designate ``Janice") jam Alice's and Bob's attempts to establish a secure quantum channel, without alerting Alice or Bob to her presence? If Janice could surreptitiously rotate the plane of polarization of the photons heading toward Alice by a large enough angle $\alpha$, then she could jam Alice's and Bob's quantum channel, without in any way affecting Alice's or Bob's instrumentation.
%(unlike other jamming attacks, such as saturating Alice's detector with a classical light source, which is readily detectable \cite{AlleaumeSurvey,HuguesSalas}). 
To accomplish such an attack, Janice could exploit Faraday rotation.

Faraday rotation arises when an external magnetic field is applied along the direction of propagation of polarized light as the light travels through a dielectric medium; under such circumstances, the plane of polarization of the light will rotate by an angle $\beta$. For light propagating a distance $L$ through a uniform medium in which a constant external field of strength $B_0$ is applied, the rotation angle $\beta$ is given by \cite{berman}
%%%%%%%%
\begin{equation}
\beta = {\cal V} B_0 L ,
\label{faraday_rotation}
\end{equation}
where ${\cal V}$, the Verdet constant, depends on properties of the medium as well as the wavelength of the light. The Verdet constant may be parameterized as \cite{berman}
%%%%%%%%
\begin{equation}
    {\cal V} = \frac{ (n^2 - 1)^2}{n} \frac{ 2 \pi^2 c \epsilon_0}{\rho e \lambda^2 } ,
    \label{Verdet}
\end{equation}
where $n$ is the index of refraction for the medium (in the absence of the external magnetic field), $c$ is the speed of light in vacuum, $\epsilon_0$ is the vacuum permittivity, $e$ is the (absolute value of the) unit charge of an electron, $\rho$ is the number density of electrons within the medium, and $\lambda$ is the wavelength of the propagating light. (Note that the index of refraction $n$ also varies weakly with $\lambda$.) As usual, we may evaluate the electron number density as
%%%%%%%%%%
\begin{equation}
    \rho = N_A \sum_i f_i y_i \frac{ \mu_i }{ m_i } ,
    \label{rho}
\end{equation}
where $N_A$ is Avogadro's number, $f_i$ is the fractional composition of the medium by constituent type $i$, $y_i$ is the number of electrons per atom or molecule of constituent type $i$, $\mu_i$ is the mass density (in ${\rm g}\, {\rm m}^{-3}$) of constituent $i$, and $m_i$ is the molecular mass (in ${\rm g} \, {\rm mol}^{-1}$) for constituent $i$. For example, the Earth's atmosphere near sea level is largely composed of $N_2$ ($f = 0.781)$, $O_2$ ($f = 0.210)$, and $Ar$ ($f = 0.01)$, for which we find $\rho = 3.7 \times 10^{26} \, {\rm m}^{-3}$.

Polarized photons traveling through the Earth's atmosphere will undergo a weak rotation, due to the Earth's magnetic field. Given the varying density of the atmosphere, one may effectively neglect such effects for altitudes $L > 4 \times 10^4 \, {\rm m}$ above sea level, and model the atmosphere as a uniform medium with fixed electron number density $\rho \simeq 3.7 \times 10^{26} \, {\rm m}^{-3}$ for $0 \leq L \leq 4 \times 10^4 \, {\rm m}$ \cite{finkel}. The Earth's magnetic field strength is also (roughly) uniform across those altitudes, with $B_0 \simeq 5 \times 10^{-5} \, {\rm T}$. For photons of wavelength $\lambda = 850$ nm, the index of refraction in air (at standard temperature and pressure) is $(n - 1) = 2.7477 \times 10^{-4}$ \cite{haynes}. Hence for the propagation of polarized photons with $\lambda = 850$ nm from space to ground \cite{yin1}, we find ${\cal V}_{\rm air} \simeq 3.7 \times 10^{-4}\, {\rm rad}\, {\rm T}^{-1} {\rm m}^{-1}$ and $\beta \simeq 7.4 \times 10^{-4}$ rad $\simeq 0.042^\circ$, in close agreement with the results for $\beta / B_0$ calculated in Ref.~\cite{finkel}. Clearly the Faraday effect arising from the Earth's own magnetic field poses no threat to successful QKD involving free-space transmission of polarized photons, including space-to-ground transmission.

For Janice to exploit the Faraday effect to jam QKD from free-space transmission of polarized photons, she would need to apply an external magnetic field considerably stronger than the Earth's field. 
%We found in Section \ref{sec:BeamRotation} that for QKD as in Ref.~\cite{yin1}, rotation by an angle $\alpha = 0.216 \> {\rm rad} = 12.4^\circ$ could effectively jam Alice's and Bob's efforts to establish a secure quantum channel. 
In order to produce a rotation $\beta \geq \alpha$ by means of the Faraday effect, Janice would need to establish $B_0 L \geq (B_0 L)_{\rm min}$, with
%%%%%%%
\begin{equation}
    (B_0 L)_{\rm min} \equiv \alpha / {\cal V} .
    \label{B0Lmin}
\end{equation}
We found in Section \ref{sec:BeamRotation} that for distribution of polarizaton-entangled photons as in Ref.~\cite{yin1}, rotation by an angle $\alpha = 0.216 \> {\rm rad} = 12.4^\circ$ could effectively jam Alice's and Bob's efforts to establish a secure quantum channel. For such free-space propagation, using ${\cal V}_{\rm air} = 3.7 \times 10^{-4} \> {\rm rad} \, {\rm T}^{-1} \, {\rm m}^{-1}$, achieving a minimum rotation angle $\alpha = 0.216 \> {\rm rad}$ would require $(B_0 L)_{\rm min} = 594.6 \> {\rm T \, m}$. 

To date, the highest continuous magnetic field strength produced by Earthbound instrumentation is $B_0 = 45\, {\rm T}$, achieved using a hybrid superconducting and resistive-materials magnet weighing 35 tons and measuring 6.7m high \cite{FSUmagnet}. Presumably Janice could not install such instrumentation somewhere along the path of Alice's photons without revealing her scheme. More typical field strengths, with $B_0 \sim {\cal O} (1) \, {\rm T}$, are produced by tiny and inexpensive objects, such as coin-sized neodymium-iron-boron rare-earth magnets \cite{Neodymium1,Neodymium2}. In this case, the challenge for Janice would be to suspend an array of such magnets over a considerable distance $(\sim 0.5$ km) along the path of Alice's photons. It therefore seems that QKD involving free-space transmission of polarized photons --- including transmission from low-Earth orbit to ground stations --- is not in jeopardy of being jammed via induced Faraday rotation.

On the other hand, most QKD systems in practice will likely depend upon  propagation through optical fiber to connect networks within and between metropolitan areas, ranging over distances from ${\cal O} (10^0 - 10^2)$ km \cite{ZeilingerTokyo,geneva,PanReviewLargeScale,PanReviewRealisticQKD}. Empirical measurements of the effective Verdet constant for typical, commercial optical fiber for wavelengths ranging across the visible bands into the near infrared may be parameterized as ${\cal V}_{\rm fiber} = a \times 10^{-28} \, \nu^2 \> {\rm rad} \, {\rm T}^{-1} \, {m}^{-1}$, with $\nu = c / \lambda$ the frequency of light in Hz. Two sets of measurements found $a = 0.159 \pm 0.008$ \cite{noda} and $a = 0.142 \pm 0.004$ \cite{cruz}. QKD via optical fiber typically uses photons with $\lambda = 1550$ nm \cite{brodsky,takesue,ZeilingerTokyo,trapateau}; adopting the lower (more conservative) measurement of ${\cal V}$ from Ref.~\cite{cruz}, this corresponds to ${\cal V}_{\rm fiber} = 0.53 \, {\rm rad} \, {\rm T}^{-1} {\rm m}^{-1}$. (We find a comparable estimate for ${\cal V}_{\rm fiber}$ if we use Eq.~(\ref{Verdet}), naively assume pure silicon dioxide for the core of the fiber, and take $n = 1.44$ for $\lambda = 1550$ nm within the fiber core.)

Given that the Verdet constant is three orders of magnitude greater for optical fiber than for the atmosphere, Janice's jamming attack based on Faraday rotation can be significantly more effective against QKD via optical fiber. In fact, the authors of Ref.~\cite{brodsky} found a weak but measurable effect on fiber-based QKD involving polarized photons arising simply from Faraday rotation from the Earth's weak magnetic field. If Janice were to apply a stronger external field $B_0$, she could achieve a sufficient rotation angle $\alpha$ to disrupt Alice's and Bob's quantum channel for $B_0 L \geq (B_0 L)_{\rm min}$, with $(B_0 L)_{\rm min} = \alpha / {\cal V}_{\rm fiber}$. 

In Tables \ref{tab:fiberQKD1} and \ref{tab:fiberQKD2} we show the requirements for Janice's jamming attack based on Bell-CHSH tests involving propagation of polarized photons with $\lambda \simeq 1550$ nm through optical fiber, both for table-top experiments (across a distance $d = 0$ km) and for long-distance propagation through fiber (across $d = 20$ km). As in Section \ref{sec:BeamRotation}, we assume that the Faraday rotation would not affect the systematic errors reported in the original experiments, and hence we calculate the shift $S_{\rm meas} \pm \sigma \rightarrow \tilde{S}_{\rm meas} \pm \sigma$, with $\tilde{S}_{\rm meas}$ related to $S_{\rm meas}$ by Eq.~(\ref{Stilde}). We consider scenarios in which Alice and Bob stop trusting the security of their quantum channel if they measure only modest violation of the Bell-CHSH inequality: either a violation by only $1.7$ standard deviations (corresponding to a probability that they were {\it not} measuring entangled particles of $p = 0.05$); or the more stringent threshold of $2.5$ standard deviations (corresponding to $p = 0.006$). Each scenario corresponds to a different minimum rotation angle $\alpha$, and hence, via Eq.~(\ref{B0Lmin}), to a different value of $(B_0 L)_{\rm min}$. 

\begin{table}[ht]
\def\arraystretch{1.2}
\begin{center}
\begin{tabular}{ |c | c| c | c| c | c|c| } 
\hline
{\rm Ref.} & $d$ [km] & $S_{\rm meas}$ & $\sigma$ & $\tilde{S}_{\rm meas}$ & $\alpha$ [rad] & $(B_0 L)_{\rm min}$ [T m]  \\
\hline
\cite{takesue} & 0 & $2.65$ & $0.09$ & $2.15$ & $0.31$ & $0.58$\\
\cite{takesue} & 20 & $2.55$ & $0.09$ & $2.15$ & $0.28$ & $0.53$\\
\hline
\cite{trapateau} & 0 & $2.57$ & $0.05$ & $2.08$ & $0.31$ & $0.58$ \\
\cite{trapateau} & 20 & $2.24$ & $0.09$ & $2.15$ & $0.14$ & $0.26$ \\
\hline
\end{tabular}
\caption{Minimum product of the field strength and path length $(B_0 L)_{\rm min}$ over which Janice would need to establish an external magnetic field, if her goal were to reduce Alice's and Bob's measured value of the CHSH parameter from $S_{\rm meas} \pm \sigma$ to $\tilde{S}_{\rm meas} \pm \sigma$ by inducing a rotation by angle $\alpha$ in the plane of polarization of the photons directed toward Alice via optical fiber. In this case, we assume that Alice and Bob set the threshold for security of their quantum channel to be violation of the Bell-CHSH inequality with $p \leq 0.05$.}
\label{tab:fiberQKD1}
\end{center}
\end{table}

\begin{table}[ht]
\def\arraystretch{1.2}
\begin{center}
\begin{tabular}{ |c | c| c | c| c | c|c| } 
\hline
{\rm Ref.} & $d$ [km] & $S_{\rm meas}$ & $\sigma$ & $\tilde{S}_{\rm meas}$ & $\alpha$ [rad] & $(B_0 L)_{\rm min}$ [T m]  \\
\hline
\cite{takesue} & 0 & $2.65$ & $0.09$ & $2.23$ & $0.29$ & $0.55$ \\
\cite{takesue} & 20 & $2.55$ & $0.09$ & $2.23$ & $0.25$ & $0.47$ \\
\hline
\cite{trapateau} & 0 & $2.57$ & $0.05$ & $2.13$ & $0.30$ & $0.57$ \\
\cite{trapateau} & 20 & $2.24$ & $0.09$ & $2.23$ & $0.05$ & $0.09$ \\
\hline
\end{tabular}
\caption{Same as Table ~\ref{tab:fiberQKD1}, but for the case in which Alice and Bob set the threshold for security of their quantum channel to be violation of the Bell-CHSH inequality with $p \leq 0.006$.}
\label{tab:fiberQKD2}
\end{center}
\end{table}

The limiting case would be if Alice and Bob could achieve $S_{\rm meas} = 2\sqrt{2}$ with vanishing systematic error. In that case, Janice would need to rotate the plane of polarization of Alice's photons by $\alpha = 0.39\> {\rm rad} = 22.5^\circ$ in order to produce $\tilde{S}_{\rm meas} = 2$. For polarized photons of $\lambda = 1550$ nm traveling through optical fiber with ${\cal V}_{\rm fiber} = 0.53 \> {\rm rad} \, {\rm T}^{-1} \, {\rm m}^{-1}$, the limit case corresponds to $(B_0 L)_{\rm min} = 0.74 \, {\rm T m}$. Like the values shown in Tables \ref{tab:fiberQKD1} and \ref{tab:fiberQKD2}, with $0.09 \leq (B_0 L)_{\rm min} \leq 0.58$, such a rotation could be readily achieved by Janice using a small number of inexpensive, coin-sized neodymium-iron-boron magnets arranged within a region of space just a few centimeters across.

In real implementations of QKD via optical fiber, transmission has been shown to remain stable over time-scales of 1 - 2 hours \cite{ZeilingerTokyo,JenneweinAlignment}. Nonetheless, recalibration mechanisms are typically employed, because tiny mechanical or thermal disturbances to the fiber can lead to significant induced birefringence, which in turn can push the polarization bases out of alignment. Typically the recalibration consists of sending (unentangled) photons of known polarization to Alice and/or Bob, who can then adjust compensating wave plates at their detector stations to realign the polarization bases. 
%When well insulated from mechanical or thermal disturbances, QKD transmission via optical fiber has been shown to be stable over time-scales of 1 - 2 hours \cite{ZeilingerTokyo,JenneweinAlignment}. 
Polarization realignment can now be implemented automatically, triggered (for example) by a rise in QBER, and typically requires $\sim 5$ s \cite{JenneweinAlignment}. Other active polarization-control methods --- which constantly monitor and adjust alignment rather than being triggered by a rise in QBER --- have been able to compensate for rapid changes in polarization basis with scrambling frequencies up to $~ 40 \pi \> {\rm rad} \, {\rm s}^{-1}$ \cite{Xavier2009,Xavier2011}. 

If Janice simply placed her magnets near the optical fiber that directed photons toward Alice's detectors and left them there, then Alice and Bob could re-establish a secure quantum channel by recalibrating their polarization bases; indeed, Alice and Bob probably would not even suspect that Janice had tried to jam their channel. However, if Janice moved her magnets in some pattern near the relevant optical fiber --- thereby changing the applied external field over a time-scale shorter than the time required to complete a polarization realignment --- then she could succeed in interrupting Alice's and Bob's quantum channel indefinitely. For the automated trigger-based realignment system of Ref.~\cite{JenneweinAlignment}, Janice would need to shift her applied field on time-scales shorter than $5$ s; for the continuous-adjustment schemes described in Refs.~\cite{Xavier2009,Xavier2011}, Janice would need to shift her applied field on time-scales shorter than $\sim 0.3 \> {\rm rad} / ( 40 \pi \>{\rm rad}\, {\rm s}^{-1} ) = 0.002$ s. Given variation of the applied magnetic field over such short time-scales, the so-called ``reference-frame-independent" QKD protocols \cite{LaingRFI} would not be sufficient to counter or withstand such a jamming attack. 

Lastly, we note that Janice could also try to accomplish her jamming attack by exploiting some effects other than Faraday rotation. For example, if she were to apply an external electric field of strength $E_0$, that would induce a shift in the effective index of refraction in the optical fiber via the Kerr effect, $\Delta n = \lambda K E_0^2$. Given typical Kerr constants $K$ for crystals \cite{haynes} and photon wavelength $\lambda = 1550$ nm, however, Janice would need to find some means of producing $E_0 \sim {\cal O} (10^9)$ V/m in order to induce a shift $\Delta n = 0.1$. Hence we believe that the magnetic effect we have analyzed here would be the most inexpensive and simple means of implementing a surreptitious jamming attack.

\section{Conclusions}
\label{sec:Conclusions}

We have identified a potential vulnerability of quantum key distribution (QKD) using polarized photons. In particular, an adversary (whom we dub Janice) could induce a rotation of the plane of polarization of the light headed toward Alice (or Bob) by applying a strong external magnetic field across a region of space through which the photons traveled. The induced Faraday rotation would prevent Alice and Bob from establishing a secure quantum channel, either by lowering the measured value of the Bell-CHSH parameter $S$ with which they would certify the security of their channel (as in the Ekert91 protocol \cite{ekert}), or, more generally, by increasing the quantum bit error rate (QBER) above acceptable levels. Such a jamming attack would not enable Janice to gain secret information (as in eavesdropping attacks), though it would enable her to disrupt Alice's and Bob's channel selectively and indefinitely without revealing her presence, and without requiring direct access to either Alice's or Bob's instrumentation or their shared optical fiber. 

Janice could further combine this jamming attack with other types of attack that might reveal unauthorized information, such as a ``device calibration" attack akin to that described in Ref.~\cite{JainDeviceCalibration}, since Janice could control the time at which Alice and/or Bob would need to recalibrate their devices. Moreover, the Faraday rotation induced by Janice should accentuate detector-efficiency mismatches at either Alice's or Bob's detectors, since, in general, detector efficiencies $\eta_0$ and $\eta_1$ for polarized photons in a given basis will not remain unchanged upon a rotation of the photons' plane of polarization. In general, Alice's and Bob's key generation rate will fall as the detector-efficiency mismatches rise \cite{FungDEM,LydersenDEM,FeiDEM}. In addition, although various QKD protocols and entanglement-verification techniques have been identified that would enable Alice and Bob to incorporate a known detector-efficiency mismatch and still preserve secure communication \cite{FungDEM,ZhangDEM,LydersenDEM,FeiDEM}, such prior calibrations would no longer accurately characterize the detector-efficiency mismatches at Alice's and/or Bob's devices upon even a modest rotation by Janice of the photons' plane of polarization. Hence the jamming attack that we describe here could, in principle, be used in conjunction with eavesdropping attacks that exploit calibration or detector-efficiency mismatches.

The magnitude of the Faraday effect depends sensitively on the medium through which the photons propagate. We find that such a jamming attack would likely not be effective against free-space transmission of the polarized photons (including space-based QKD as in Ref.~\cite{LiaoSatellite,YinSatelliteQKD,MiciusQKD}), though it could be more readily effective against QKD systems that depend upon transmission of polarized photons through optical fiber. In particular, given the sensitivity of optical fiber to mechanical disturbances, installations of QKD fiber networks will likely shield against direct mechanical interference, which could impede Janice's ability to intervene surreptitiously with the fibers (for example, by blocking or cutting them). Given cost considerations, however, it is unlikely that tens or even hundreds of kilometers of optical fiber for QKD will be protected by robust magnetic shielding. Hence Janice could implement her jamming attack without gaining direct, physical access to the fibers, while remaining far away from either Alice's or Bob's receiving stations. 

Given the easy availability of inexpensive rare-earth, coin-sized magnets with field strengths $B_0 \sim {\cal O} (1)\> {\rm T}$ (available at local hardware stores), Janice's jamming attack could be both inexpensive and inconspicuous. A few such magnets arranged across a few centimeters (concealed, say, within a backpack) and placed near an unguarded stretch of optical fiber could disrupt Alice's and Bob's quantum channel. A simple mechanical construction within the backpack, which would rearrange the magnets in space on an appropriately short time-scale and thereby shift the direction and strength of the applied field in an unpredictable pattern, would spoil Alice's and Bob's efforts to reestablish a secure channel via polarization realignment or recalibration. 

The jamming attack we have identified here targets QKD involving polarized photons. Whether comparable attacks could be effective against other popular QKD protocols involving photons, such as time-energy entanglement, remains the subject of further research.

\section*{Acknowledgements}

We gratefully acknowledge helpful discussions with Jason Gallicchio, Calvin Leung, Thomas Scheidl, and William Wootters. Portions of this work were conducted in MIT's Center for Theoretical Physics and supported in part by the U.S. Department of Energy under Contract No. DE-SC0012567, as well U.S. Department of Energy Contract No. DE-SC0011091. DK also acknowledges support from NSF INSPIRE Award PHY-1541160.

%\bibliography{bibliography}


\end{document}